\documentclass[final,1p,times]{elsarticle}

\usepackage{amssymb}
\usepackage{amsmath}

\begin{document}

\begin{frontmatter}



\title{A short note on reconstruction variables
in shock capturing schemes for magnetohydrodynamics}


\author[label1]{Takahiro Miyoshi\corref{cor1}}
\ead{miyoshi@sci.hiroshima-u.ac.jp}

\author[label2]{Takashi Minoshima}

\address[label1]{Graduate School of Advanced Science and Engineering,
Hiroshima University, Higashihiroshima 739-8526, Japan}

\address[label2]{Center for Mathematical Science and Advanced Technology,
Japan Agency for Marine-Earth Science and Technology, Yokohama 236-0001, Japan}

\cortext[cor1]{Corresponding author}

\begin{abstract}
We propose a set of quick and easy approximate characteristic variables
for higher-order reconstructions of shock capturing schemes
for magnetohydrodynamics (MHD).
Numerical experiments suggest that
the reconstructions using the approximate characteristic variables
are more robust than those using the conservative or primitive variables,
while their computational efficiencies are comparable.
The approximate characteristic variables are simple compared to
the full characteristic variables for MHD,
and can be a practical choice of reconstruction variables.
\end{abstract}


\end{frontmatter}


\def\vec#1{\mbox{\boldmath $#1$}}

\section{Introduction}

High-order shock capturing schemes such as
monotonic upstream-centered scheme
for conservation laws (MUSCL) \cite{vanleer79},
essentially non-oscillatory scheme (ENO) \cite{harten87},
weighted essentially non-oscillatory scheme (WENO) \cite{liu94},
and monotonicity-preserving scheme (MP) \cite{suresha97}
are designed based on the theory of hyperbolic conservation laws.
The reconstruction techniques used in the aforementioned schemes
extend a monotone first-order scheme to higher orders,
while suppressing numerical oscillations near discontinuities.
For a scalar conservation law,
the variable to be reconstructed is the conservative variable,
which is identical to the characteristic variable.
By contrast, reconstruction variables
in a system of nonlinear hyperbolic conservation laws
are not unique
because the conservative variables and characteristic ones
are different in general.

In order to avoid spurious numerical oscillations,
the reconstruction should be performed
in the characteristic fields \cite{harten87}.
However, in the magnetohydrodynamic (MHD) system,
the characteristic variables are much more complicated
than those in the hydrodynamic system \cite{brio88,roe96},
and as a result,
the reconstruction of those demands high computational cost.
For example, in an experiment using an open source MHD simulation code
CANS+ \cite{matsumoto19}, where the fifth-order MP of
the characteristic variables is adopted as the reconstruction,
the computational time of the reconstruction steps
is more than $60 \%$ of the total time.
To reduce the computational time,
a particular high-order multidimensional scheme has been proposed
that reduces the number of appearances of the characteristic variables
while maintaining robustness \cite{balsara09}.
Meanwhile, the conservative variables or the primitive variables
have often been used instead of the characteristic variables
due to computational efficiency in many practical applications.
In this paper, we propose a set of quick and easy reconstruction variables
for high-order MHD schemes,
where the reconstructions using the proposed variables are
more robust than those using the conservative or primitive variables,
while their efficiencies are comparable.

\section{Set of variables of the MHD equations}

\subsection{Conservative variables}

Consider the one-dimensional conservation laws:
\begin{equation}
\frac{\partial \vec{U}}{\partial t} + \frac{\partial \vec{F}}{\partial x} = 0,
\label{eq:cons}
\end{equation}
where $\vec{U}$ and $\vec{F}$ are
the state vector of the conservative variables and corresponding flux vector.
The set of the conservative variables of MHD is given by
\begin{equation}
\vec{U} = \left( \rho , m_x , m_y , m_z , B_y , B_z , e \right)^T,
\end{equation}
where $\rho$, $m$, $B$, and $e$ are the density, momentum, magnetic field,
and total energy density, respectively.
The subscripts $x$, $y$, and $z$ denote the $x$-, $y$-, and $z$-components
of vector fields.
In one-dimension, $B_x$ is constant in space and time
due to the solenoidal condition of the magnetic field.
The flux vector of the ideal MHD equations is
\begin{equation}
\vec{F} =
\begin{pmatrix}
m_x \\
\frac{m_x^2}{\rho} + p + \frac{B^2}{2} - B_x^2 \\
\frac{m_x}{\rho} m_y - B_x B_y \\
\frac{m_x}{\rho} m_z - B_x B_z \\
\frac{m_x}{\rho} B_y - B_x \frac{m_y}{\rho} \\
\frac{m_x}{\rho} B_z - B_x \frac{m_z}{\rho} \\
\frac{m_x}{\rho} (e + p + B^2/2) - B_x \frac{(m_x B_x + m_y B_y + m_z B_z)}{\rho}
\end{pmatrix}.
\end{equation}
When an ideal gas equation of state is considered,
the pressure $p$ is determined from
\begin{equation}
p = ( \gamma - 1 ) \left( e - \frac{m^2}{2\rho} - \frac{B^2}{2} \right),
\end{equation}
where $\gamma$ denotes the ratio of specific heats.

\subsection{Primitive variables}

The conservation laws of MHD can be rewritten as
a quasilinear form for primitive variables $\vec{V}$ \cite{powell99}:
\begin{equation}
\frac{\partial \vec{V}}{\partial t} + \vec{A}_P \frac{\partial \vec{V}}{\partial x} = 0,
\label{eq:prim}
\end{equation}
where $\vec{A}_P$ is a coefficient matrix.
An often used set of the primitive variables is given by
\begin{equation}
\vec{V} = \left( \rho , v_x , v_y , v_z , B_y , B_z , p \right)^T,
\end{equation}
where $v = m/\rho$ is the velocity.
The temperature or the entropy density can become
an alternative to the pressure.
Using the transformation matrix (see (19) and (20) in \cite{powell99}),
infinitesimal variations of the conservative and primitive variables
are related by
\begin{equation}
d \vec{V} = \vec{M}^{-1} d \vec{U}, \quad
\vec{M} = \frac{\partial \vec{U}}{\partial \vec{V}}.
\end{equation}

\subsection{Characteristic variables}

Since the ideal MHD equations are hyperbolic,
the equations are decomposed into a system of nonlinear advection equations as
\begin{equation}
\vec{R}_P^{-1} \frac{\partial \vec{V}}{\partial t}
+ \vec{R}_P^{-1} \vec{A}_P \vec{R}_P \vec{R}_P^{-1}
\frac{\partial \vec{V}}{\partial x} \equiv
\frac{\partial \vec{W}}{\partial t}
+ \vec{\Lambda} \frac{\partial \vec{W}}{\partial x} = 0,
\label{eq:char}
\end{equation}
where $\vec{R}_P$ and $\vec{\Lambda}$ are
right eigenvectors of $\vec{A}_P$
and diagonal matrix of the eigenvalues of $\vec{A}_P$, respectively.
The characteristic variables $\vec{W}$ are transformed into the others by
\begin{equation}
d \vec{W} = \vec{R}_P^{-1} d \vec{V} = \vec{R}^{-1} d \vec{U},
\end{equation}
where $\vec{R}= \vec{M} \vec{R}_P$.
The eigenvectors must be appropriately normalized
so as to be well-behaved \cite{brio88,roe96}
because the eigenvectors can be singular when the eigenvalues are degenerate.
The transformation to and from the characteristic variables
of the MHD equations is time-consuming compared with
that of the Euler equations and can be a computational bottleneck.

\subsection{Approximate characteristic variables}

Let us consider approximate subsystems reduced from the MHD equations.
The compressible subsystem is obtained by taking the limit of
$B_x \rightarrow 0$ in (\ref{eq:cons}) or (\ref{eq:prim})
to remove the magnetic tension terms.
Combining the evolution equations for $B_y$, $B_z$, and $p$,
we find
\begin{equation}
\frac{\partial}{\partial t}
\begin{pmatrix}
\rho \\
v_x \\
p^{\prime}_T
\end{pmatrix}
+
\begin{pmatrix}
v_x & \rho & 0 \\
0 & v_x & \frac{1}{\rho} \\
0 & \rho {c^{\prime}_f}^2 & v_x
\end{pmatrix}
\frac{\partial}{\partial x}
\begin{pmatrix}
\rho \\
v_x \\
p^{\prime}_T
\end{pmatrix}
= 0,
\label{eq:comp}
\end{equation}
where
\begin{equation}
p^{\prime}_T = p + \frac{1}{2} \left( B_y^2 + B_z^2 \right), \quad
\rho {c^{\prime}_f}^2 = \gamma p + B_x^2 + B_y^2 + B_z^2.
\end{equation}
The second term on the right side of the last equation
has been added to include a contribution of $B_x$.
The incompressible subsystems, on the other hand, are derived
in the limit of $\partial v_x / \partial x \rightarrow 0$ by
\begin{equation}
\frac{\partial}{\partial t}
\begin{pmatrix}
v_{y,z} \\
B_{y,z}
\end{pmatrix}
+
\begin{pmatrix}
v_x & - \frac{1}{\rho} B_x \\
-B_x & v_x
\end{pmatrix}
\frac{\partial}{\partial x}
\begin{pmatrix}
v_{y,z} \\
B_{y,z}
\end{pmatrix}
= 0.
\label{eq:incomp}
\end{equation}
The subsystems (\ref{eq:comp}) and (\ref{eq:incomp})
readily lead to the relations between the characteristic variables
of the subsystems, $W_m^{\prime}$, and the primitive variables:
\begin{equation}
\begin{pmatrix}
d W^{\prime}_1 \\
d W^{\prime}_2 \\
d W^{\prime}_3
\end{pmatrix}
=
\begin{pmatrix}
{c^{\prime}_f}^2 & 0 & -1 \\
0 & \rho c^{\prime}_f & 1 \\
0 & - \rho c^{\prime}_f & 1
\end{pmatrix}
\begin{pmatrix}
d \rho \\
d v_x \\
d p^{\prime}_T
\end{pmatrix}, \quad
\begin{pmatrix}
d W^{\prime}_{4,6} \\
d W^{\prime}_{5,7}
\end{pmatrix}
=
\begin{pmatrix}
\sqrt{\rho} & 1 \\
\sqrt{\rho} & -1
\end{pmatrix}
\begin{pmatrix}
d v_{y,z} \\
d B_{y,z}
\end{pmatrix}.
\label{eq:v2w}
\end{equation}
Meanwhile, the inverse transformations are given by
\begin{equation}
\begin{pmatrix}
d \rho \\
d v_x \\
d p^{\prime}_T
\end{pmatrix}
=
\begin{pmatrix}
\frac{1}{{c^{\prime}_f}^2} & \frac{1}{2 {c^{\prime}_f}^2} &
\frac{1}{2 {c^{\prime}_f}^2} \\
0 & \frac{1}{2 \rho c^{\prime}_f} & -\frac{1}{2 \rho c^{\prime}_f} \\
0 & \frac{1}{2} & \frac{1}{2}
\end{pmatrix}
\begin{pmatrix}
d W^{\prime}_1 \\
d W^{\prime}_2 \\
d W^{\prime}_3
\end{pmatrix}, \quad
\begin{pmatrix}
d v_{y,z} \\
d B_{y,z}
\end{pmatrix}
=
\begin{pmatrix}
\frac{1}{2 \sqrt{\rho}} & \frac{1}{2 \sqrt{\rho}} \\
\frac{1}{2} & -\frac{1}{2}
\end{pmatrix}
\begin{pmatrix}
d W^{\prime}_{4,6} \\
d W^{\prime}_{5,7}
\end{pmatrix}.
\label{eq:w2v}
\end{equation}
Using the relation $dp^{\prime}_T = dp + B_y dB_y + B_z dB_z$,
\begin{equation}
dp
= \frac{1}{2} \left(
dW^{\prime}_2 + dW^{\prime}_3
- B_y dW^{\prime}_4 + B_y dW^{\prime}_5
- B_z dW^{\prime}_6 + B_z dW^{\prime}_7
\right).
\label{eq:dp}
\end{equation}
Here $W_m^{\prime}$ is referred to as approximate characteristic variables.
In vector form, the approximate characteristic variables are related
to the conservative and primitive variables by
\begin{equation}
d \vec{W}^{\prime}
= {\vec{R}^{\prime}_P}^{-1} d \vec{V}
= {\vec{R}^{\prime}}^{-1} d \vec{U}
\end{equation}
where $\vec{R}^{\prime} = \vec{M} \vec{R}^{\prime}_P$.
The transformation matrices
${\vec{R}^{\prime}_P}^{-1}$ consisting of (\ref{eq:v2w})
and $\vec{R}^{\prime}_P$ of (\ref{eq:w2v})
are considerably simplified compared to the eigenvectors
of the complete MHD system.

\section{Reconstruction procedures}

In the finite volume approach, higher-order spatial accuracy
can be achieved by reconstructing the cell-averaged value
and evaluating the values on the left and right faces of the cell.
Similarly, in the finite difference approach,
higher-order accuracy is realized by reconstructing the point value
and evaluating the values on the left and right sides of the midpoint.

Consider first the scalar conservation law for $u$.
Let $u_{j+1/2}^L$ denote $u$ on the left
of the cell face or the midpoint, $j+1/2$.
In general, introducing a well-controlled interpolation function
$\mathcal{I}$ to suppress numerical oscillations,
\begin{equation}
u_{j+1/2}^L = \mathcal{I}
\left( \cdots , u_{j-1} , u_{j} , u_{j+1} , \cdots \right),
\label{eq:ul}
\end{equation}
where $u_{j+s}$ is the cell-averaged or the point value
at $j+s$.
Similarly, the right value $u_{j-1/2}^R$ is obtained
by reversing the order of $u_{j+s}$ in $\mathcal{I}$.

Correspondingly, we evaluate the variables of the MHD on $j+1/2$
by employing the general interpolation function $\mathcal{I}$
as the following subsections.

\subsection{Conservative variable reconstruction}

Compute the conservative variables on the left
$\vec{U}_{j+1/2}^L$ using $\vec{U}_{j+s}$:
\begin{equation}
\vec{U}_{j+1/2}^L = \mathcal{I}
\left( \cdots , \vec{U}_{j-1} , \vec{U}_{j} , \vec{U}_{j+1} , \cdots \right),
\label{eq:uurec}
\end{equation}
where $\mathcal{I}$ acts on each component of the vector.
Hereinafter,
(\ref{eq:uurec}) is referred to as $\vec{U}$-{\it reconstruction}.

\subsection{Primitive variable reconstruction}

Compute the primitive variables $\vec{V}_{j+1/2}^L$
using $\vec{V}_{j+s} \equiv \vec{V} ( \vec{U}_{j+s} )$ as
\begin{equation}
\vec{V}_{j+1/2}^L = \mathcal{I}
\left( \cdots , \vec{V}_{j-1} , \vec{V}_{j} , \vec{V}_{j+1} , \cdots \right).
\label{eq:vvrec}
\end{equation}
Hereinafter referred to as $\vec{V}$-{\it reconstruction}.

\subsection{Characteristic variable reconstruction}

Compute $\vec{V}_{j+1/2}^L$
by interpolating the characteristic variables
applied with the eigenvectors for the primitive system at $j$ as
\cite{balsara98,balsara04}
\begin{equation}
\vec{V}_{j+1/2}^L = {\vec{R}_P}_j \mathcal{I}
\left( \cdots , \bar{\vec{W}}_{j-1} , \bar{\vec{W}}_{j} ,
\bar{\vec{W}}_{j+1} , \cdots \right),
\quad
\bar{\vec{W}}_{j+s} = {\vec{R}_P}_j^{-1} \vec{V}_{j+s}.
\label{eq:vwrec}
\end{equation}
Hereinafter referred to as $\vec{W}$-{\it reconstruction}.
Note that, for example in the finite-difference WENO schemes
\cite{jiang99,balsara00},
the characteristic decomposition is performed
using averaged eigenvectors at the midpoint $j+1/2$.
This paper, however, does not focus on that.
Also, $\vec{U}_{j+1/2}^L$ can be directly computed by applying $\vec{R}$:
\begin{equation}
\vec{U}_{j+1/2}^L = \vec{R}_j \mathcal{I}
\left( \cdots , \tilde{\vec{W}}_{j-1} , \tilde{\vec{W}}_{j} ,
\tilde{\vec{W}}_{j+1} , \cdots \right),
\quad
\tilde{\vec{W}}_{j+s} = \vec{R}_j^{-1} \vec{U}_{j+s}.
\label{eq:uwrec}
\end{equation}
Although $\vec{U}_{j+1/2}^L \neq \vec{U} (\vec{V}_{j+1/2}^L)$ in general,
both lead to similar results
in later tests (but not shown).

\subsection{Approximate characteristic variable reconstruction}

Compute $\vec{V}_{j+1/2}^L$ applying the transformation matrix
$\vec{R}^{\prime}_P$ to the approximate characteristic variables
$\vec{W}^{\prime}$
(referred to as $\vec{W}^{\prime}$-{\it reconstruction}):
\begin{equation}
\vec{V}_{j+1/2}^L = {\vec{R}^{\prime}_P}_j \mathcal{I}
\left( \cdots , \bar{\vec{W}^{\prime}}_{j-1} , \bar{\vec{W}^{\prime}}_{j} ,
\bar{\vec{W}^{\prime}}_{j+1} , \cdots \right),
\quad
\bar{\vec{W}^{\prime}}_{j+s}
= {{\vec{R}^{\prime}_P}_j}^{-1} \vec{V}^{\prime}_{j+s}.
\label{eq:varec}
\end{equation}
Alternatively,
compute $\vec{U}_{j+1/2}^L$ applying the transformation matrix
$\vec{R}^{\prime}$ to the approximate characteristic variables
$\vec{W}^{\prime}$ (results not shown):
\begin{equation}
\vec{U}_{j+1/2}^L = \vec{R}^{\prime}_j \mathcal{I}
\left( \cdots , \tilde{\vec{W}^{\prime}}_{j-1} , \tilde{\vec{W}^{\prime}}_{j} ,
\tilde{\vec{W}^{\prime}}_{j+1} , \cdots \right),
\quad
\tilde{\vec{W}^{\prime}}_{j+s}
= {\vec{R}^{\prime}_j}^{-1} \vec{U}^{\prime}_{j+s}.
\label{eq:uarec}
\end{equation}

\section{Numerical experiments}

Typical numerical experiments are presented.
The Harten-Lax-van Leer Discontinuities (HLLD) approximate
Riemann solver \cite{miyoshi05} is employed as a base scheme for MHD.
The second-order MUSCL and
the fifth-order weighted compact nonlinear scheme (WCNS),
which is constructed by combining the fourth-order interpolation
and fourth-order finite difference \cite{minoshima19},
are adopted as high-order schemes.
The third-order TVD Runge-Kutta method \cite{shu88}
is used for the time integration in all the experiments.

\begin{figure}
\centering
\includegraphics[scale=0.8]{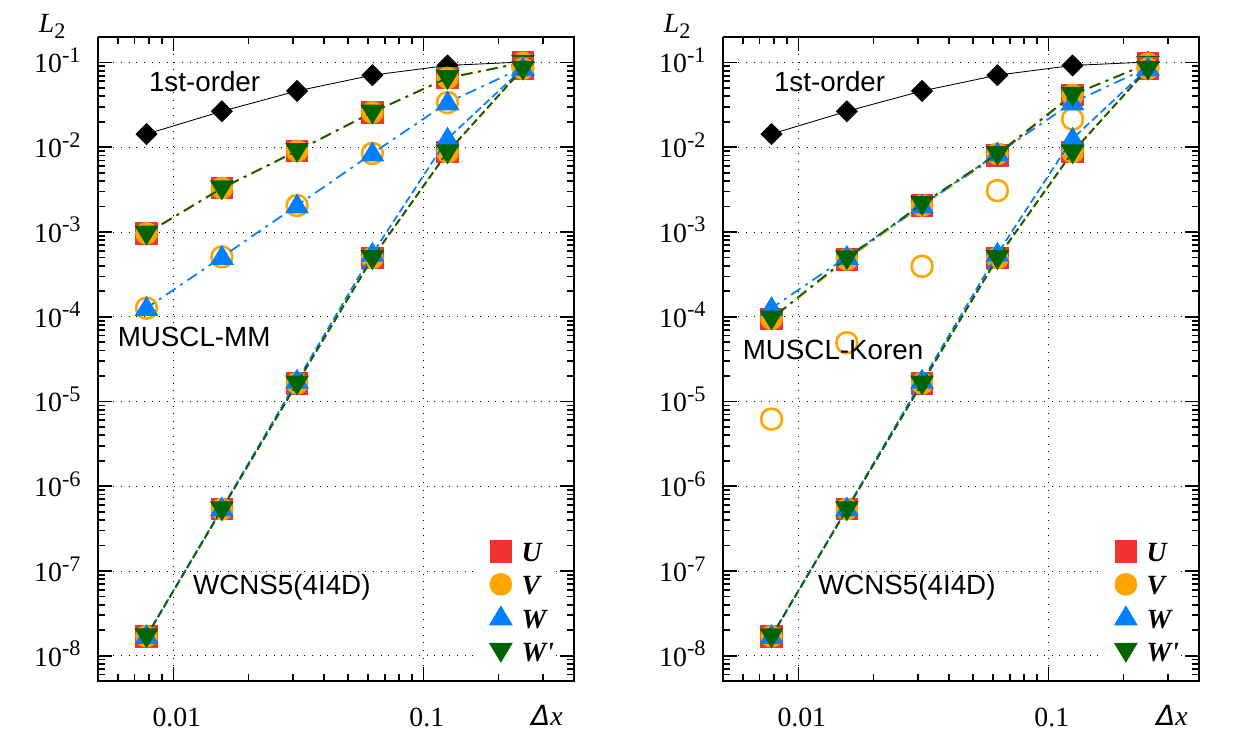}
\caption{
The $L_2$-norm errors of the magnetic field strength
for different grid resolutions.
Each color corresponds to a different {\it reconstruction}.
(Left)
The base scheme (solid line),
MUSCL scheme with the minmod limiter (dash-dotted lines),
and WCNS scheme (dashed lines).
Orange open circles show the second-order linear reconstruction.
(Right)
The MUSCL scheme with the Koren limiter (dash-dotted lines).
The base and WCNS schemes are also included because of visibility.
Orange open circles show the third-order linear reconstruction.
\label{fig:fig1}}
\end{figure}

In Fig.~\ref{fig:fig1},
the accuracy of the high-order schemes
with the different reconstruction variables is examined by
performing the one-dimensional circularly polarized Alfv\'{e}n wave test
\cite{toth00}.
The time step size is fixed small enough
so as not to be affected by errors of the time integral.
The CFL number at the highest resolution is on the order of $10^{-4}$.
The results indicate that $\vec{W}$-{\it reconstruction}
in the MUSCL scheme with the minmod limiter is more accurate than
$\vec{U}$-, $\vec{V}$-, and $\vec{W}^{\prime}$-{\it reconstructions}
because the limiter is not activated
in the distribution of the exact characteristic variables.
On the other hand, the higher-order reconstructions,
including the MUSCL scheme with the Koren limiter \cite{koren93},
do not depend on the reconstruction variables
and achieve the expected order of accuracy.
Note that although the Koren limiter can reconstruct
a smooth distribution with third-order accuracy,
it is limited to second-order in this problem.

\begin{figure}
\centering
\includegraphics[scale=0.8]{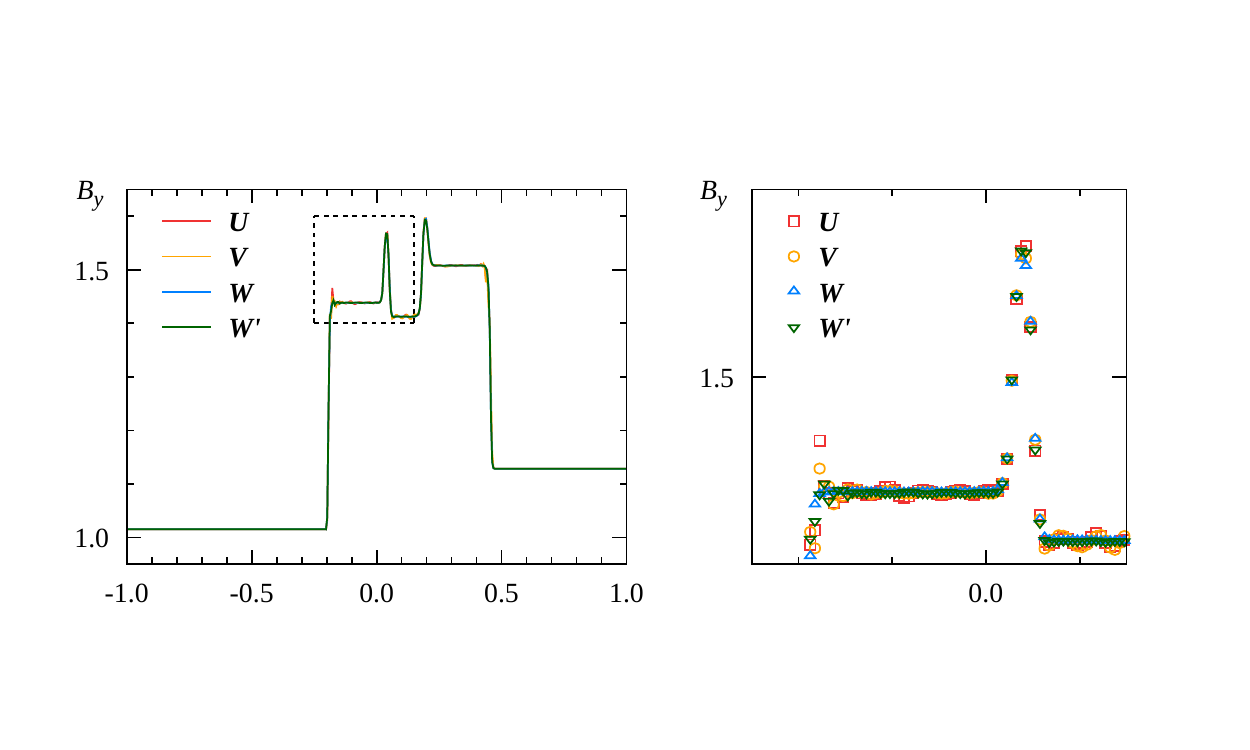}
\caption{
Dai-Woodward shock tube problem computed by the WCNS scheme.
$B_y$ is plotted with the color corresponding to
each set of reconstruction variables.
The right is an enlarged view.
\label{fig:fig2}}
\end{figure}

\begin{figure}
\centering
\includegraphics[scale=0.8]{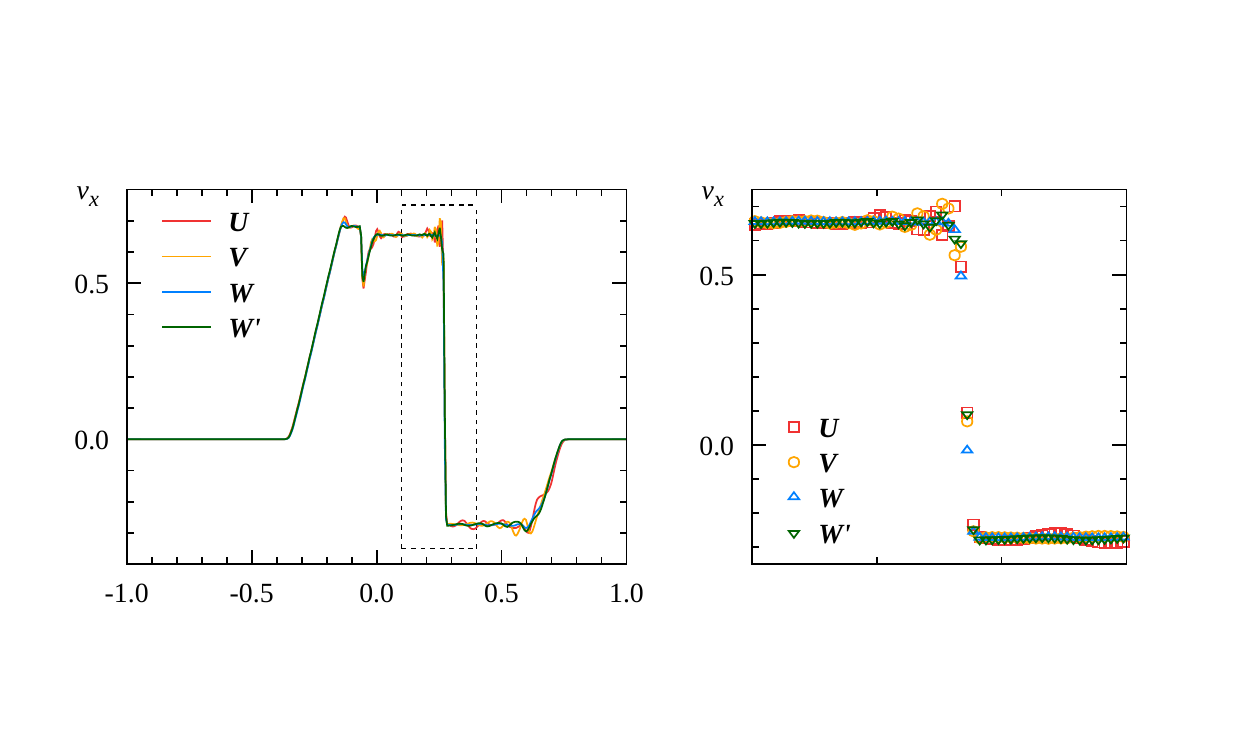}
\caption{
Brio-Wu shock tube problem computed by the WCNS scheme.
$V_x$ is plotted with the color corresponding to
each set of reconstruction variables.
The right is an enlarged view.
\label{fig:fig3}}
\end{figure}

We present a typical one-dimensional shock tube problem \cite{dai94},
which contains two fast shocks, two slow shocks,
two rotational discontinuities, and one contact discontinuity.
The left and right states are initialized as
$\vec{V}^L$ $=$ $(1.08,1.2,0.01,0.5,3.6/\sqrt{4 \pi},2/\sqrt{4 \pi},0.95)^T$
and $\vec{V}^R$ $=$ $(1,0,0,0,4/\sqrt{4 \pi},2/\sqrt{4 \pi},1)^T$
with $B_x = 2/\sqrt{4 \pi}$, respectively.
The CFL number is $0.8$ in this test.
Using the MUSCL scheme with the minmod limiter,
all {\it reconstructions} produce the similar solutions
without numerical oscillations.
The results obtained by the WCNS scheme, however,
reveal that considerable overshoots are observed
in $\vec{U}$- and $\vec{V}$-{\it reconstructions}
as shown in Fig.~\ref{fig:fig2}.
By contrast, the WCNS scheme with $\vec{W}^{\prime}$-{\it reconstruction}
is as robust as that with $\vec{W}$-{\it reconstruction}
though small bumps are observed.
Another typical shock tube problem \cite{brio88} is also demonstrated.
This problem contains a number of different waves:
two fast rarefaction waves, one slow shock, one slow compound wave,
and a contact discontinuity.
The left and right states are initially given by
$\vec{V}^L$ $=$ $(1,0,0,0,1,0,1)^T$
and $\vec{V}^R$ $=$ $(0.125,0,0,0,-1,0,0.1)^T$
with $B_x = 0.75$.
The MUSCL-minmod scheme with each {\it reconstruction}
gives similar reasonable results.
On the other hand, the WCNS scheme with each {\it reconstruction}
yields slightly different solutions as shown in Fig.~\ref{fig:fig3}.
Large numerical oscillations are observed
in $\vec{U}$- and $\vec{V}$-{\it reconstructions}.
In $\vec{W}^{\prime}$-{\it reconstruction}, however,
numerical oscillations are reduced
though those are still larger than in $\vec{W}$-{\it reconstruction}.
Note that a numerical oscillation with a relatively long wavelength
attached to the right-moving fast rarefaction wave is observed
even in $\vec{W}$-{\it reconstruction}.
Furthermore, we performed several shock tube problems consisting of
pure waves such as isolated shocks and rarefaction waves \cite{falle98},
and confirmed that the results for $\vec{W}^{\prime}$-{\it reconstruction}
are very similar to those for $\vec{W}$-{\it reconstruction} (not shown).

\begin{figure}
\centering
\includegraphics[scale=1.0]{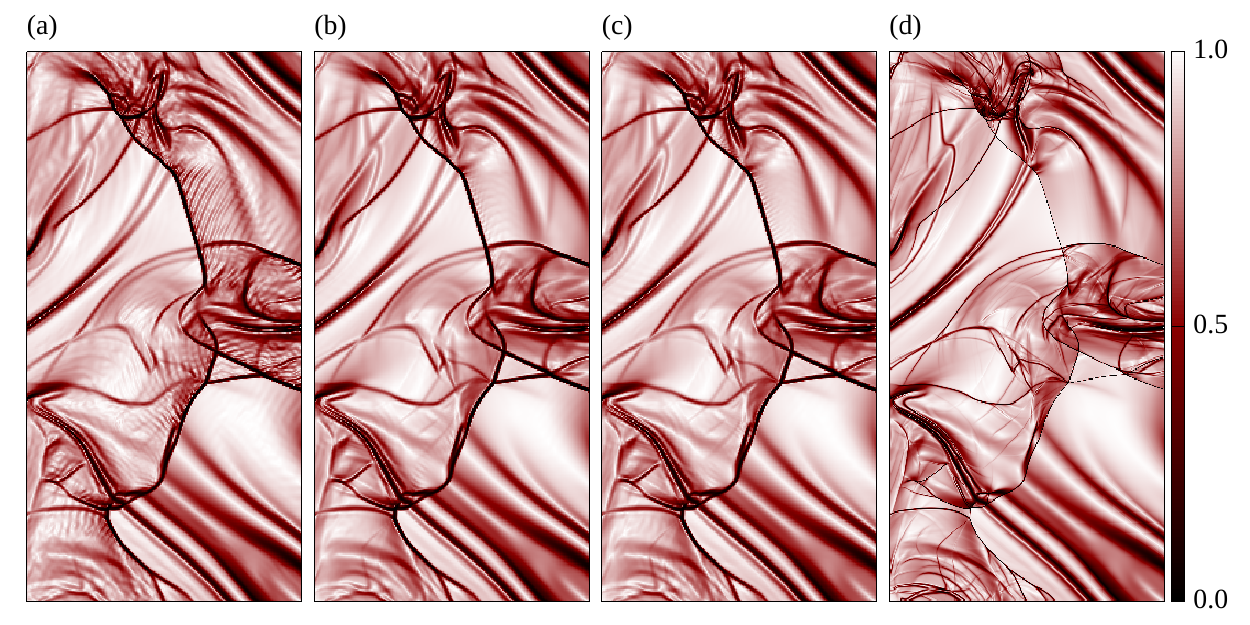}
\caption{
Schlieren-like images calculated as
$\exp \left( - 10 | \nabla \rho | / | \nabla \rho |_{\max} \right)$
for the Orszag-Tang vortex obtained by the WCNS scheme with
(a) $\vec{V}$-{\it reconstruction}, (b) $\vec{W}$-{\it reconstruction},
and (c) $\vec{W}^{\prime}$-{\it reconstruction}.
(d) The reference solution obtained by the MUSCL-minmod scheme with
$\vec{W}$-{\it reconstruction} on $4000^2$ grid.
Since the resolution of the grid is $10$ times higher,
$\exp \left( - 100 | \nabla \rho | / | \nabla \rho |_{\max} \right)$ is shown
for comparison.
\label{fig:fig4}}
\end{figure}

Finally, let us demonstrate the Orszag-Tang vortex problem
\cite{orszag79}, which is a standard test for multidimensional MHD.
Considering the trade-off between computational cost and accuracy,
the high-order schemes are necessary, especially
for practical multidimensional problems.
To maintain the solenoidal condition of the magnetic field,
the hyperbolic divergence cleaning method \cite{dedner02} is used.
The number of the grid is $400^2$ for $x,y \in [0,2 \pi]$,
and the CFL number is set to $0.4$.
Fig.~\ref{fig:fig4} displays schlieren-like images
that visualize density variations, computed by the WCNS scheme.
The WCNS scheme with $\vec{W}$- and $\vec{W}^{\prime}$-{\it reconstructions}
yields a similar result to the reference solution.
In contrast to that,
visible numerical oscillations attached to shocks
are observed in high-compression regions
when $\vec{U}$- (not shown) and $\vec{V}$-{\it reconstructions} are adopted.

\section{Conclusions}

We have proposed a set of quick and easy reconstruction variables
for high-order shock capturing schemes for MHD.
The numerical experiments suggest that
$\vec{W}^{\prime}$-{\it reconstruction} is capable of
the higher-order reconstructions almost without numerical oscillations
even when $B_x \neq 0$ or $d v_x / dx \neq 0$.
In the present WCNS code,
the computational time of $\vec{W}$-{\it reconstruction}
is more than $3.5$ times ($3.6$ times in the MUSCL-minmod code)
larger than that of $\vec{U}$- and $\vec{V}$-{\it reconstructions},
while the time of $\vec{W}^{\prime}$-{\it reconstruction}
is about $1.1$ times ($1.4$ times in the MUSCL-minmod code).
As a result,
the high-order schemes with $\vec{W}^{\prime}$-{\it reconstruction}
are more robust than those with $\vec{U}$- and $\vec{V}$-{\it reconstructions},
while their computational efficiencies are comparable.
Thus we conclude that $\vec{W}^{\prime}$-{\it reconstruction}
can be a practical choice for the higher-order reconstructions
in shock capturing schemes for MHD.

In terms of robustness or accuracy,
$\vec{W}^{\prime}$-{\it reconstruction} can be switched to
$\vec{W}$-{\it reconstruction} or others adaptively
in a very problematic situation.
Indeed, another set of reconstruction variables
that improves the accuracy in low Mach number MHD flows
has been proposed \cite{minoshima20}.
We expect that the present approach
of introducing reasonably approximated characteristic fields
can be extended to more complex systems.
The system of relativistic MHD may be an example
since the corresponding eigensystem is too complicated
to be used for the higher-order reconstructions.

\section*{Acknowledgement}

This work was supported by MEXT/JSPS KAKENHI Grant Numbers
JP20K11851, JP20H00156, JP19H01928 (T. Miyoshi).




\begin{thebibliography}{00}


\bibitem{balsara98}
D. Balsara,
Total variation diminishing scheme for adiabatic and isothermal magnetohydrodynamics,
Astrophys. J. Suppl. 116 (1998) 133-153.

\bibitem{balsara04}
D. Balsara,
Second-order-accurate schemes for magnetohydrodynamics with divergence-free reconstruction,
Astrophys. J. Suppl. 151 (2004) 149-184.

\bibitem{balsara00}
D. Balsara, C.~W. Shu,
Monotonicity preserving weighted essentially non-oscillatory schemes with increasingly high order of accuracy,
J. Comput. Phys. 160 (2000) 405-452.

\bibitem{balsara09}
D. Balsara, T. Rumpf, M. Dumbser, C.~D. Munz,
Efficient, high accuracy ADER-WENO schemes for hydrodynamics and divergence-free magnetohydrodynamics,
J. Comput. Phys. 228 (2009) 2480-2516.

\bibitem{brio88}
M. Brio, C.~C. Wu,
An upwind differencing scheme for the equations of ideal magnetohydrodynamics,
J. Comput. Phys. 75 (1988) 400-422.

\bibitem{dai94}
W. Dai, P.~R. Woodward,
An approximate Riemann solver for ideal magnetohydrodynamics,
J. Comput. Phys. 111 (1994) 354-372.

\bibitem{dedner02}
A. Dedner, F. Kemm, D. Kr\"{o}ner, C.~D. Munz, T. Schnitzer, M. Wesenberg,
Hyperbolic divergence cleaning for the MHD equations,
J. Comput. Phys. 175 (2002) 645-673.

\bibitem{falle98}
S.~A.~E.~G. Falle, S.~S. Komissarov, P. Joarder,
A multidimensional upwind scheme for magnetohydrodynamics,
Mon. Not. R. Astron. Soc. 297 (1998) 265-277.

\bibitem{harten87}
A. Harten, B. Engquist, S. Osher, S.~R. Chakravarthy,
Uniformly high order accurate essentially non-oscillatory schemes, III,
J. Comput. Phys. 71 (1987) 231-303.

\bibitem{jiang99}
G.~S. Jiang, C.~C. Wu,
A high-order WENO finite difference scheme for the equations of ideal magnetohydrodynamics,
J. Comput. Phys. 150 (1999) 561-594.

\bibitem{koren93}
B. Koren,
A robust upwind discretization method for advection,
diffusion and source terms,
in: C.~B. Vreugdenhil, B. Koren (Eds.),
Numerical Methods for Advection–Diffusion Problems,
Vieweg, Braunschweig, Germany, 1993, pp. 117-138.

\bibitem{liu94}
X.~D. Liu, S. Osher, T. Chan,
Weighted essentially non-oscillatory schemes,
J. Comput. Phys. 115 (1994) 200-212.

\bibitem{matsumoto19}
Y. Matsumoto, Y. Asahina, Y. Kudoh, T. Kawashima, J. Matsumoto,
H.~R. Takahashi, T. Minoshima, S. Zenitani, T. Miyoshi, R. Matsumoto,
Magnetohydrodynamic simulation code CANS+: Assessments and applications,
Publ. Astron. Soc. Japan 71 (2019) 83(1-26).

\bibitem{minoshima19}
T. Minoshima, T. Miyoshi, Y. Matsumoto,
A high-order weighted finite difference scheme
with a multistate approximate Riemann solver
for divergence-free magnetohydrodynamic simulations,
Astrophys. J. 242 (2019) 14(1-29).

\bibitem{minoshima20}
T. Minoshima, K. Kitamura, T. Miyoshi,
A multistate low-dissipation advection upstream splitting method
for ideal magnetohydrodynamics,
Astrophys. J. 248 (2020) 12(1-21).

\bibitem{miyoshi05}
T. Miyoshi, K. Kusano,
A multi-state HLL approximate Riemann solver for ideal magnetohydrodynamics,
J. Comput. Phys. 208 (2005) 315-344.

\bibitem{orszag79}
A. Orszag, C.~M. Tang,
Small-scale structure of two-dimensional magnetohydrodynamic turbulence,
J. Fluid. Mech. 90 (1979) 129-143.

\bibitem{powell99}
K.~G. Powell, P.~L. Roe, T.~J. Linde, T.~I. Gombosi, D.~L. De Zeeuw,
A solution-adaptive upwind scheme for ideal magnetohydrodynamics,
J. Comput. Phys. 154 (1999) 284-309. 

\bibitem{roe96}
P.~L. Roe, D.~S. Balsara,
Notes on the eigensystem of magnetohydrodynamics,
SIAM J. Appl. Math. 56 (1996) 57-67.

\bibitem{shu88}
C.~W. Shu, S. Osher,
Efficient implementation of essentially non-oscillatory shock-capturing schemes,
J. Comput. Phys. 77 (1988) 439-471.

\bibitem{suresha97}
A. Suresha, H.~T. Huynh,
Accurate monotonicity-preserving schemes with Runge–Kutta time stepping,
J. Comput. Phys. 136 (1997) 83-99.

\bibitem{toth00}
G. T\'{o}th,
The $\nabla \cdot \vec{B}$ constraint
in shock-capturing magnetohydrodynamic codes,
J. Comput. Phys. 161 (2000) 605-652.

\bibitem{vanleer79}
B. van Leer,
Towards the ultimate conservative difference scheme, V.
A second order sequel to Godunov's method,
J. Comput. Phys. 32 (1979) 101-136.

\end{thebibliography}


\clearpage

\clearpage

\clearpage

\clearpage

\end{document}